\documentclass[reprint,aps,pra, 10pt,
superscriptaddress,
amsmath,amssymb,
floatfix
]{revtex4-1} 

\usepackage{amsmath,amssymb,bm}
\usepackage{array}
\usepackage{booktabs}
\usepackage{esvect}
\usepackage{graphicx}
\usepackage{mathtools}
\usepackage{multirow}
\usepackage{comment}
\usepackage{hyperref}

\usepackage[svgnames]{xcolor}

\begin{document}

\title{Do Atoms Age?}

\author{Mark G. Raizen}

\affiliation{Department of Physics, The University of Texas at Austin, Austin, TX 78712}

\author{David E. Kaplan}
\affiliation{Department of Physics \& Astronomy, The Johns Hopkins University, Baltimore, MD  21218, USA}

\author{Surjeet Rajendran}
\affiliation{Department of Physics \& Astronomy, The Johns Hopkins University, Baltimore, MD  21218, USA}

\date{\today}

\begin{abstract}
Time evolution generically entangles a quantum state with environmental degrees of freedom. The resulting increase in entropy changes the properties of that quantum system leading to ``aging''. It is interesting to ask if this familiar property also applies to simple, single particle quantum systems such as the decay of a radioactive particle. We propose a test of such aging in an ion clock setup where we probe for temporal changes to the energies of the electronic state of an ion containing a radioactive nucleus. Such effects are absent in standard quantum mechanics and this test is thus a potent null test for violations of quantum mechanics. As a proof of principle, we show that these effects exist in causal non-linear modifications of quantum mechanics. 
\end{abstract}

\maketitle


{\it Introduction}.--- The Schrodinger equation in quantum mechanics is time-symmetric, with no past and no future.  This successful description of the microscopic world is in sharp contrast to the macroscopic world where there is a clear “arrow of time”. In  quantum mechanics, the ``arrow of time'' is an emergent phenomenon that arises due to the entanglement between a quantum system and its environment, with the associated increase in the overall entropy providing a ``direction'' to the ``arrow of time''. This view is in fact necessary to define non-trivial quantum cosmological dynamics \cite{Page:1982fk} since unlike quantum mechanics where time is regarded as a privileged marker of dynamical evolution, General Relativity treats it simply as a parameter. Generically, the entanglement between a complex quantum system and the environment results in changes to the physical properties of the quantum system {\it i.e. } the quantum system {\it ages}. It is interesting to ask if such a concept of aging could also exist in simple quantum systems such as single particle dynamics. 

There is a natural place for such a notion to manifest itself in single particle dynamics - namely, the decay of an unstable quantum system wherein the system gets entangled with a large reservoir of final states. In standard quantum mechanics, when an unstable quantum system is created, the decay process decreases the probability of finding the quantum system in that initial state - however, whenever the system is found in its initial state, its properties are unchanged {\it i.e.} the entanglement of the unstable system to the reservoir of final states does not change the properties of the system itself. An experimental test of this phenomenon thus becomes a powerful probe of new physics beyond standard quantum mechanics. While quantum mechanics remains unchanged since it was developed almost 100 years ago, there are no known fundamental reasons why it should resist parametrized deviation. Concrete and logically consistent models that modify the time evolution of quantum mechanics have been written down. This includes generalizations of unitary Hamiltonian evolution to non-unitary but probability preserving Lindblad evolution \cite{Weinberg:2016uml} as well as causal, unitary but non-linear Hamiltonian evolution \cite{Kibble:1978vm, Stamp:2015vxa,  Kaplan:2021qpv}. 

We propose a test of such aging using radioactive atoms  with a half-life that is long enough to probe it repeatedly as an atomic clock.   Could a clock transition in such an atom have a time-dependent frequency shift as it ages?  If so, such radioactive atoms could be distinguished from each other by their age, a concept that is analogous to our macroscopic world.  Such experimental searches can already be performed, as outlined below for several candidate systems including lutetium and radium. We also show that this phenomenon of atom aging arises naturally in the non-linear quantum mechanical framework developed in \cite{Kaplan:2021qpv}. This experiment is thus a  test of this particular framework, but it is possible that the described phenomenon may arise generically in other modifications of quantum mechanics as well. 

The rest of this paper is organized as follows. We next describe the experimental protocol and discuss the specific application of this protocol. We then estimate the effects of the non-linear quantum evolution \cite{Kaplan:2021qpv} for this class of experiments, and make concluding remarks.

{\it Setup}.--- The general protocol proposed here is the following. The first step is to create a radioisotope with an appropriate half-life ranging from several days to weeks. Typically, one starts with a stable precursor isotope which is enriched to a high level by existing methods.  Nuclear transmutation is most commonly accomplished by neutron irradiation in a nuclear reactor or by proton bombardment in a cyclotron, though other accelerator methods can be used. After transmutation, the radioisotope can be isolated from the stable target by radiochemistry \cite{Choppin:2002} or by physical separation methods \cite{Mazur:2014}. It would then be shipped to a metrology lab where the radioisotopes could be evaporated and trapped as neutral atoms or ions. The clock transition in the atoms would be repeatedly measured over a time scale comparable to the half-life of the radioisotope. A change in the transition energy would be a signal of atom aging. 

\begin{figure}[h!]
\centering
\includegraphics[width=2.5in]{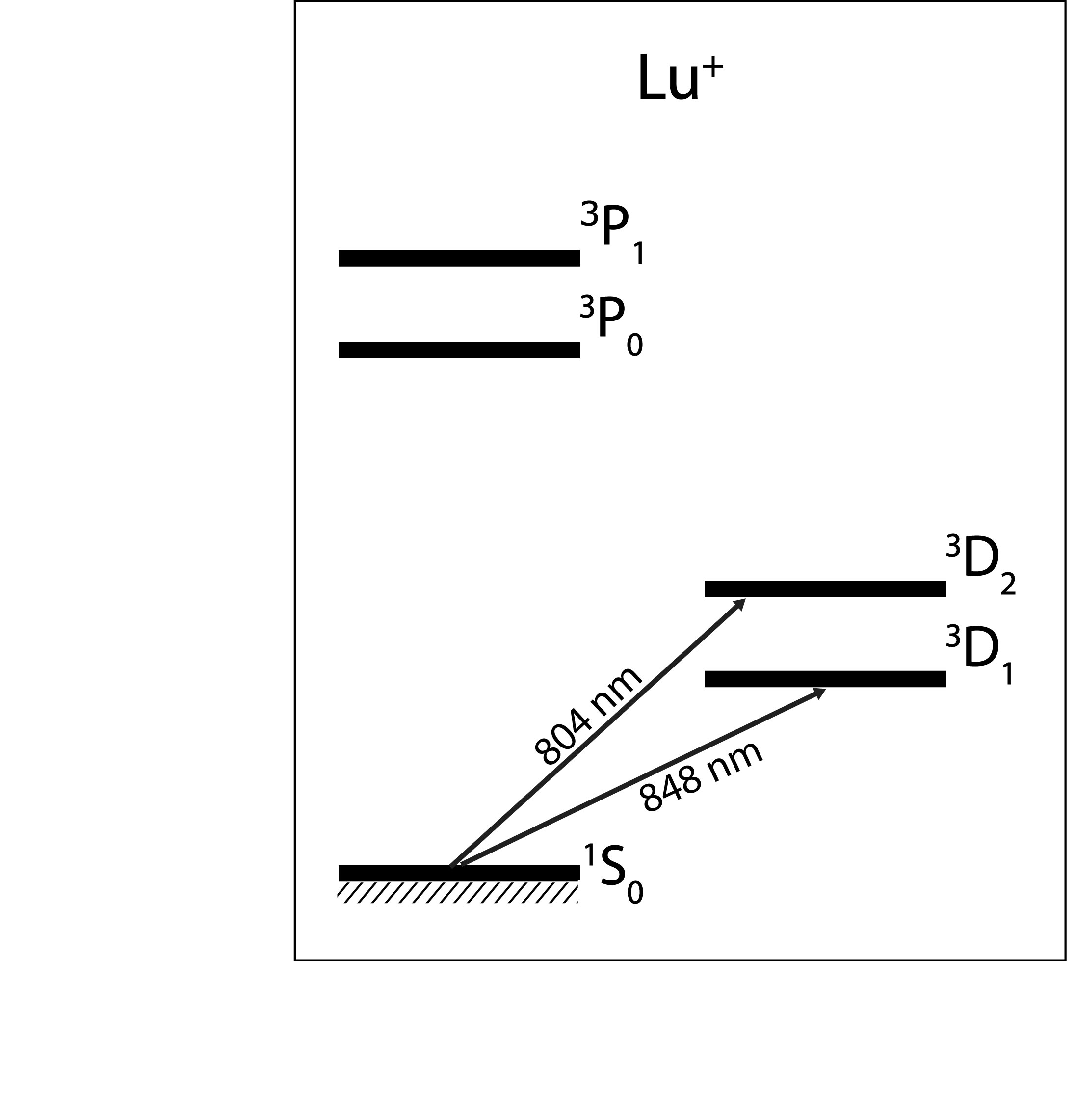}
\caption{The level structure of Lu$^{+}$ with two possible clock transitions.}
\label{figLu}
\end{figure}

{\it Lutetium}.--- The first case we consider is Lutetium (Lu), a lanthanide with two stable isotopes.  A promising candidate is $^{177}$Lu$^{+}$ with a half-life of 6.65 days, decaying by beta emission to stable $^{177}$Hf.  This radioisotope is produced by irradiation of stable $^{176}$Yb in a nuclear reactor, which is transmuted to $^{177}$Yb by neutron capture, decaying to $^{177}$Lu with a half-life of 1.9 hours.  The radioisotope $^{177}$Lu can be efficiently separated from the target $^{176}$Yb by radiochemistry or physical separation \cite{Mazur:2014}.  These methods produce no-carrier-added $^{177}$Lu which is the most promising radioisotope today for targeted cancer therapy \cite{Banerjee:2015}.  Clock transitions in Lu$^{+}$ have been measured, and the black-body shift was found to be the lowest of any established optical clocks \cite{Arnold:2018}. A simplified schematic is shown in Fig. \ref{figLu}.

There are two possible clock transitions in Lu$^{+}$, an electric quadrupole transition (E2) near 804 nm, and a magnetic dipole transition (M1) near 848 nm.  The latter clock transition can be detected with a second laser near 646 nm.  Optical pumping of the ground state is accomplished with two lasers near 622 nm and 350 nm.  These measurements were performed with $^{176}$Lu$^{+}$, but could similarly be performed with $^{177}$Lu$^{+}$.  The hyperfine splitting will be different, due to the different nuclear spins (7 for $^{176}$Lu and 7/2 for $^{177}$Lu) as well as different magnetic moments.  A better comparison is stable $^{175}$Lu which has a very high natural abundance and same nuclear properties as $^{177}$Lu. Alternative ion clocks with radioisotopes are $^{169}$Er or $^{175}$Yb with half-lives of 9.375 days and 4.185 days respectively.


{\it Radium}.--- We now consider  Radium-223 ($^{223}$Ra). This is a pure alpha emitter, decaying to $^{219}$Rn with a half-life of 11.43 days. $^{223}$Ra is actually used in medicine as $^{223}$Ra-Chloride to treat metastatic prostate cancer that has spread to the bone marrow. It is produced in the decay chain of $^{227}$Ac, which in turn is produced by proton spallation of $^{232}$Th or by neutron irradiation of $^{226}$Ra.
\begin{figure}[h!]
\centering
\includegraphics[width=2.5in]{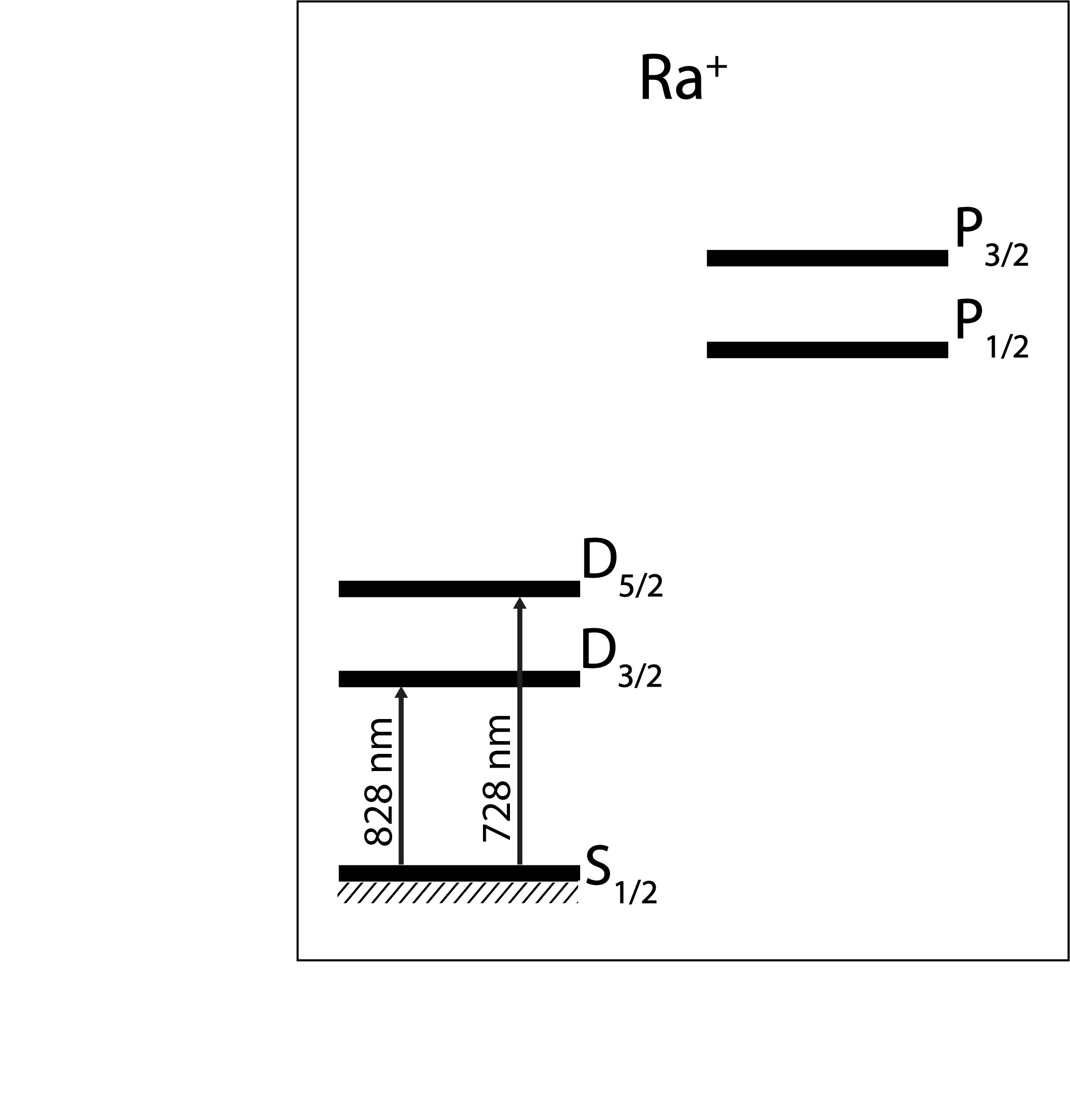}
\caption{The level structure of Ra$^{+}$ with two possible clock transitions.}
\label{figRa}
\end{figure}

Experiments to date have been performed on singly-ionized $^{226}$Ra which was confined in a RF Paul trap and spectroscopic measurements were reported in a series of papers \cite{Jayich:2019, Jayich:2020}.  A schematic showing the energy levels and transition wavelengths is shown in Fig. \ref{figRa}.  These are electric quadrupole transitions at 728 nm and 828 nm. 

{\it Theory}.--- In this section, as an example of a theory where atom aging is possible, we compute the effects of the non-linear quantum mechanical evolution proposed in \cite{Kaplan:2021qpv} on the electronic state of an ion.  This modification preserves causality, energy conservation, and gauge-invariance of the theory, as well as ensuring that quantum states have a conserved norm. A detailed description of  \cite{Kaplan:2021qpv} is beyond the scope of this paper and thus we will summarize the physics that is of direct relevance to this experiment.  In this theory, the non-linear evolution of a quantum state $|\chi\rangle$ is constructed on top of the linear quantum mechanical interactions that already exist in the theory. In this paper, we consider effects arising from electromagnetism and QCD. While it will turn out that ion clocks are not the best tests of this non-linear modification, the intent of this section is to show that the concept of atom aging can arise in a consistent modification of quantum mechanics.


{\it Electromagnetism}.--- To incorporate non-linear evolution, in the path integral of electromagnetism,   the interaction $e A_{\mu} J^{\mu}$ between the electromagnetic field $A_{\mu}$ and a current $J^{\mu}$ is modified to: $e A_{\mu} J^{\mu} \rightarrow e \left(A_{\mu}+\epsilon_{\gamma}\langle\chi|A_{\mu}|\chi\rangle\right) J^{\mu}$,
where $\epsilon_{\gamma}$ is the parameter that quantifies the degree of non-linearity for electromagnetic fields and $e$ is the electric charge. In the perturbative description of this theory, the leading order  ({\it i.e.} $\mathcal{O}\left(\epsilon_{\gamma}\right)$) effects of the non-linearity are computed by treating $\langle\chi|A_{\mu}|\chi\rangle$ as a background classical field in the path integral after gauge fixing, with $\langle\chi|A_{\mu}|\chi\rangle$ computed at the zeroth order {\it i.e.} using the linear quantum evolution of the theory. 

This modification to the path integral implies that the time evolution of an ion in the quantum state $|\chi\rangle$ is: 

\begin{equation}
    i \frac{\partial |\chi\rangle}{\partial t} = \left( H_{L} + \epsilon_{\gamma} \, e\,  \langle \chi |A_{0}| \chi \rangle \right)  |\chi\rangle
    \label{Eqn:NL}
\end{equation}
where $H_L$ is the usual linear quantum mechanical Hamiltonian and $\langle \chi |A_{0}| \chi \rangle$ is the expectation value of the Coulomb potential ($A_0$) of the electromagnetic field in the full quantum state $|\chi\rangle$ of the ion.  

A key aspect of \eqref{Eqn:NL} is that the state $|\chi\rangle$ represents the full quantum state of the ion, both its electronic and center of mass degrees of freedom. This is unlike the case of linear quantum mechanics where the time evolution of the electronic state of an ion can be decoupled from the evolution of its center of mass.  Such a separation is not possible in non-linear quantum mechanics. 

For the purpose of estimating this effect, we idealize the experimental setup by considering an ion that is trapped within a size $l$. We estimate the effect perturbatively in $\epsilon_{\gamma}$ - thus, to estimate the non-linear effects to $\mathcal{O}\left(\epsilon_{\gamma}\right)$, it is sufficient to know the quantum state $|\chi_0\rangle$ to zeroth order in $\epsilon_{\gamma}$ {\it i.e.} in standard quantum mechanics. This state is:  

\begin{equation}
    |\chi_0\left(t\right)\rangle = e^{-\frac{\Gamma t}{2}} |\psi_T\rangle +  \sqrt{1 - e^{-\Gamma t}} |\psi_F\rangle
\end{equation}
Here $\Gamma$ is the decay rate of the ion, $|\psi_T\rangle$ the full wave-function of the trapped ion and $|\psi_F\rangle$ the wave-function of the daughter ion.  We work in the limit where the daughter ion is not trapped in the potential. In this limit, the daughter ion freely leaves the trap and thus its contribution to the potential in the trap is negligible. The Coulomb potential $\langle \chi_0 |A_0|\chi_0\rangle$ is: 

\begin{equation}
    \langle \chi_0 |A_0|\chi_0\rangle \sim \frac{e^{-\Gamma t} \epsilon_{\gamma} e}{4 \pi l}
\end{equation}

The electron in the ion thus experiences this time dependent potential which becomes exponentially small over time $t \gg \frac{1}{\Gamma}$. This potential acts as a background electric field 

\begin{equation}
    E_{NL} \sim \frac{e^{-\Gamma t} \epsilon_{\gamma} e}{4 \pi l^2}
    \label{Eqn:NLElec}
\end{equation}

which shifts the energy levels of the ion via a second order Stark shift. This second order energy shift $\Delta E$ is of order: 

\begin{equation}
    \Delta E \approxeq E^2_{NL} \alpha_P  
    \label{Eqn:NLShift}
\end{equation}
where $\alpha_P$ is the polarizability of the ionic state. Since the norm of the state $|\chi_0\rangle$ becomes exponentially small after a time $t \gtrapprox \frac{1}{\Gamma}$, the electric field  in \eqref{Eqn:NLElec} and the associated energy shift in \eqref{Eqn:NLShift} will vanish after a time $t \gtrapprox \frac{1}{\Gamma}$. Thus, by comparing the electronic energy when the ion was first produced to the energy after time $t \gtrapprox \frac{1}{\Gamma}$, we may look for evidence of atom aging produced as a result of non-linear quantum evolution. 

For this energy shift to be visible in an ion clock setup, the energy shift \eqref{Eqn:NLShift} must give rise to a phase shift $\delta \phi$ between different electronic states of the ion. This phase shift is: 

\begin{equation}
    \delta \phi \approxeq E^2_{NL} \Delta \alpha_P T
\end{equation}
where $\Delta \alpha_P$ is the difference in the polarizability of the different electronic states of the ion and $T$ is the interrogation time of the experiment.

Ion clock states are specifically chosen so that $\Delta \alpha_P$ is small, in order to minimize backgrounds from stray electric fields and blackbody clock shifts. Since the non-linear electromagnetic effect also appears as a background electric field $E_{NL}$, the effect of these non-linearities in ion clock setups are significantly suppressed by design. 

As a result of this suppression by design, this setup is not a particularly potent probe of the electromagnetic effects described in  \cite{Kaplan:2021qpv}. One may estimate the value of $\epsilon_{\gamma}$ that can be probed by this setup by comparing the value of the non-linearly produced electric field $E_{NL}$ with typical blackbody electric fields $\sim$ kV/m, yielding $\epsilon_{\gamma} \gtrapprox 10^{-5} \left(\frac{l}{\text{10 nm}} \right)^2$. While not as compelling as other probes of  \cite{Kaplan:2021qpv}, this potential limit on $\epsilon_{\gamma}$ is still many orders of magnitude better than the model independent limit from measurements of the Lamb shift on this parameter. It also serves as a proof of concept of how the phenomenon of ``atom aging'' can arise in a logically consistent theory. 


{\it QCD}.--- The experiments proposed in this paper are at low energies. It is thus appropriate to first describe non-linear quantum mechanical terms using the low energy degrees of freedom of QCD. We comment on the UV completion at the end of this section. At low energies, the relevant QCD degrees of freedom are nucleons and mesons such as the pion and the rho-meson. To incorporate non-linear evolution, in the path integral of nuclear physics, the interaction $g \pi \bar{\Psi}\Psi$ between the neutral pion $\pi$ and a nucleon $\Psi$ gets modified to (for example): $g \pi \bar{\Psi}\Psi \rightarrow g \left(\pi + \epsilon_{N} \langle \chi | \pi | \chi \rangle \right)  \bar{\Psi}\Psi$ where $\epsilon_N$ is the strength of the non-linearity in the nucleon and $g$ the pion-nucleon scalar coupling.

In the presence of this non-linear term, a nucleus spread over a distance $l$ in a trap sources a classical pion field $\pi_{cl}$: 
\begin{equation}
    \pi_{cl} \sim \frac{g A \epsilon_N}{l^3 m_{\pi}^2}
\end{equation}
where $m_{\pi}$ is the mass of the pion and $A$ the atomic number. This classical field shifts the mass of a nucleon  by $\Delta M$ where
\begin{equation}
\Delta M =  g \pi_{cl} \sim \frac{g^2 A^2 \epsilon_N}{l^3 m_{\pi}^2}.
\end{equation} 
Similar to the derivation of \eqref{Eqn:NLElec}, this mass shift is also time dependent when the nucleus undergoes decay, proving that the concept of atom aging can be extended to nuclear decays as well. 

How can this shift to the nucleon mass be measured? In the ion-clock setup, a shift to the mass of a nucleon will appear as an isotopic correction to the energy levels. Due to the suppressed nature of the isotopic correction ($\sim$ GHz), it can be verified that with current ($\sim$ mHz) sensitivities to energy shifts, these effects are not visible in ion clock setups. This effect is also suppressed in canonical Mossbauer setups since these involve isomeric transitions that by design result in the daughter nucleus existing in the same lattice state as the parent. Thus, there isn't a significant change to the classical pion field $\pi_{cl}$ as a result of the decay. However, the effect is significant in nuclear decays where the daughter nucleus gets kicked out of the spatial location of the parent -  but, these small mass changes are difficult to detect in such systems. 

The best system to observe this particular effect is likely to be ion (or possibly, atom) interferometers where the ion is placed in a spatial superposition. In such a setup, the nuclear effects considered here will shift the mass of the nucleus in an arm by an amount proportional to the intensity (or contrast) of the arm, leading to an intensity dependent phase shift in the experiment. We leave a detailed analysis of this measurement for future work. 

 We now sketch how the low energy non-linear QCD interaction described above can arise from the high energy theory. At high energies, the degrees of freedom are quarks $q$ and gluons $A^{a}_{\mu}$. To incorporate the non-linearity, we modify the path integral of QCD to change the interaction term $g_{3}A^{a}_{\mu} \bar{q} \gamma^{\mu} q \rightarrow g_{3} A^{g}_{\mu} \left( \bar{q} \gamma^{\mu} T^a q + \epsilon_{g} \langle \chi | \bar{q} \gamma^{\mu} T^a q |\chi\rangle  \right)$ where $g_3$ is the gauge coupling of QCD and $\epsilon_g$ is the strength of the non-linearity in QCD. It can be checked that the path integral thus obtained is gauge invariant.  For simplicity, we have taken $\epsilon_g$ to be flavor universal. A nucleon is a color neutral object and thus to produce the effective nucleon-pion coupling described above, it requires a two gluon insertion, one of which is sourced by the non-linear terms.

{\it Conclusions}.--- In this paper, we have described a setup to search for the aging of an atom where the time evolution of an unstable atom causes its properties to change over time. This effect is absent in quantum mechanics and this proposal is thus a null test to search for deviations from quantum mechanics. We showed that such effects are present in non-linear modifications to quantum mechanics such as the framework described in \cite{Kaplan:2021qpv}. 

While this phenomenon is present in \cite{Kaplan:2021qpv}, one drawback of the proposed test of atom aging for electromagnetism is that ion clocks are designed to suppress background electromagnetic fields, which is the signal of the non-linearity. Non-linear effects tied to QCD can shift the mass of the nucleus and these can change the properties of decaying nuclear states, thus showing that the concept of atom aging can also be extended to nuclear physics. Beyond the specific setup of \cite{Kaplan:2021qpv}, it is interesting to ask if the concept of atom aging manifests itself in other frameworks that have attempted to modify quantum mechanics such as the more general Lindlad evolution suggested in \cite{Weinberg:2016uml} or in other approaches \cite{Weinberg:2014ewa, Griffiths:1993zz, Cotler:2022weg}.

{\it Acknowledgements}.--- We would like to thank Dima Budker for discussions, and Scott Bustabad for assistance with the manuscript.  S.R. and D.K. are supported in part by the U.S.~National Science Foundation (NSF) under Grant No.~PHY-1818899.   
This work was supported by the U.S.~Department of Energy (DOE), Office of Science, National Quantum Information Science Research Centers, Superconducting Quantum Materials and Systems Center (SQMS) under contract No.~DE-AC02-07CH11359. 
S.R.~is also supported by the DOE under a QuantISED grant for the MAGIS Collaboration using the resources of the Fermi National Accelerator Laboratory (Fermilab), and the Simons Investigator Award No.~827042.


\begin{thebibliography}{0}%
\makeatletter
\providecommand \@ifxundefined [1]{%
 \@ifx{#1\undefined}
}%
\providecommand \@ifnum [1]{%
 \ifnum #1\expandafter \@firstoftwo
 \else \expandafter \@secondoftwo
 \fi
}%
\providecommand \@ifx [1]{%
 \ifx #1\expandafter \@firstoftwo
 \else \expandafter \@secondoftwo
 \fi
}%
\providecommand \natexlab [1]{#1}%
\providecommand \enquote  [1]{``#1''}%
\providecommand \bibnamefont  [1]{#1}%
\providecommand \bibfnamefont [1]{#1}%
\providecommand \citenamefont [1]{#1}%
\providecommand \href@noop [0]{\@secondoftwo}%
\providecommand \href [0]{\begingroup \@sanitize@url \@href}%
\providecommand \@href[1]{\@@startlink{#1}\@@href}%
\providecommand \@@href[1]{\endgroup#1\@@endlink}%
\providecommand \@sanitize@url [0]{\catcode `\\12\catcode `\$12\catcode
  `\&12\catcode `\#12\catcode `\^12\catcode `\_12\catcode `\%12\relax}%
\providecommand \@@startlink[1]{}%
\providecommand \@@endlink[0]{}%
\providecommand \url  [0]{\begingroup\@sanitize@url \@url }%
\providecommand \@url [1]{\endgroup\@href {#1}{\urlprefix }}%
\providecommand \urlprefix  [0]{URL }%
\providecommand \Eprint [0]{\href }%
\providecommand \doibase [0]{http://dx.doi.org/}%
\providecommand \selectlanguage [0]{\@gobble}%
\providecommand \bibinfo  [0]{\@secondoftwo}%
\providecommand \bibfield  [0]{\@secondoftwo}%
\providecommand \translation [1]{[#1]}%
\providecommand \BibitemOpen [0]{}%
\providecommand \bibitemStop [0]{}%
\providecommand \bibitemNoStop [0]{.\EOS\space}%
\providecommand \EOS [0]{\spacefactor3000\relax}%
\providecommand \BibitemShut  [1]{\csname bibitem#1\endcsname}%
\let\auto@bib@innerbib\@empty
\end{thebibliography}%


\begin{thebibliography}{10}
\expandafter\ifx\csname url\endcsname\relax
  \def\url#1{{\tt #1}}\fi
\expandafter\ifx\csname urlprefix\endcsname\relax\def\urlprefix{URL }\fi

\bibitem{Page:1982fk}
D.~N.~Page,
Gen. Rel. Grav. \textbf{14}, 299-302 (1982)
doi:10.1007/BF00756064.

\bibitem{Weinberg:2016uml}
S.~Weinberg,
Phys. Rev. A \textbf{94}, no.4, 042117 (2016).
doi:10.1103/PhysRevA.94.042117
[arXiv:1610.02537 [quant-ph]].

\bibitem{Kibble:1978vm}
T.~W.~B.~Kibble,
Commun. Math. Phys. \textbf{64}, 73-82 (1978)
doi:10.1007/BF01940762

\bibitem{Stamp:2015vxa}
P.~C.~E.~Stamp,
New J. Phys. \textbf{17}, no.6, 065017 (2015)
doi:10.1088/1367-2630/17/6/065017
[arXiv:1506.05065 [gr-qc]].

\bibitem{Kaplan:2021qpv}
D.~E.~Kaplan and S.~Rajendran,
[arXiv:2106.10576 [hep-th]].

\bibitem{Choppin:2002}
G.~Choppin, J.-O.~Liljenzin, and J.~Rydberg, {\it Radiochemistry and nuclear chemistry}. Butterworth-Heinemann, 2002.

\bibitem{Mazur:2014}
T.~R.~~Mazur, B.~G.~Klappauf and M.~G.~Raizen, Nature Physics {\bf 10}, 601 (2014) 

\bibitem{Banerjee:2015}
S.~Banerjee, M.~R.~A.~Pillai, and F.~F.~Knapp, Chemical Reviews {\bf 115}, 2934 (2015)


\bibitem{Kozlov:2013}
A.~Kozlov, V.~A.~Dzuba and V.~V.~Flambaum, Phys. Rev. A {\bf 88}, 032509 (2013)

\bibitem{Hoyt:2005}
C.~W.~Hoyt, Z.~W.~Barber, C.~W.~Oates, T.~M.~Fortier, S.~A.~Diddams and L.~Hollberg, Phys. Rev. Lett {\bf 95}, 083003 (2005) 

\bibitem{Godun:2014}
R.~M.~Godun, {\it et. al.}, Phys. Rev. Lett. {\bf 113}, 210801 (2014) 
\bibitem{Huntemann:2016}
N.~Huntemann, C.~Sanner, P.~Lipphardt, Chr.~ Tamm, and E.~Peik, Phys. Rev. Lett. {\bf 116}, 063001 (2016) 

\bibitem{Arnold:2018}
K.~J.~Arnold, R.~Kaewuam, A.~Roy, T.~R.~Tan and M.~D.~Barrett, Nature Communications {\bf 9}, 1650 (2018)

\bibitem{Jayich:2019}
C.~A.~Holliman, M.~Fan and A.~M.~Jayich, Phys. Rev. A. \textbf{100}, 062512 (2019)

\bibitem{Jayich:2020}
C.~A.~Holliman, M.~Fan, A.~Contractor, M.~W.~Strauss and A.~M.~Jayich, Phys. Rev. A. \textbf{102}, 042822 (2020)





\bibitem{Weinberg:2014ewa}
S.~Weinberg,
Phys. Rev. A \textbf{90}, no.4, 042102 (2014)
doi:10.1103/PhysRevA.90.042102
[arXiv:1405.3483 [quant-ph]].



\bibitem{Griffiths:1993zz}
R.~B.~Griffiths,
Phys. Rev. Lett. \textbf{70}, 2201-2204 (1993)
doi:10.1103/PhysRevLett.70.2201

\bibitem{Cotler:2022weg}
J.~Cotler and A.~Strominger,
[arXiv:2201.11658 [hep-th]].

\end{thebibliography}
\end{document}